\documentclass[fleqn,usenatbib]{mnras}

\usepackage{newtxtext,newtxmath}

\usepackage[T1]{fontenc}
\usepackage{ae,aecompl}
\usepackage{xspace}

\usepackage[markup=underlined]{changes}
\colorlet{Changes@Color}{red}

\newcommand{\vesc}{\ensuremath{v_{\rm esc}}}
\newcommand{\xltwo}{\ensuremath{x_{\rm L2}}}
\newcommand{\ei}{\ensuremath{E_{\rm i}}}
\newcommand{\ef}{\ensuremath{E_{\rm f}}}
\newcommand{\ltwo}{L2\xspace}
\newcommand{\msun}{\ensuremath{M_\odot}}

\usepackage{graphicx}	
\usepackage{amsmath}	
\usepackage{amssymb}	
\usepackage{physics}

\title[Mass loss from L2 point]{Kinematics of Mass Loss from the Outer Lagrange Point L2}

\author[D. Hubov\'a and O. Pejcha]{Dominika Hubov\'a\thanks{Contact e-mail: \href{mailto:dominika.hubova@gmail.com}{dominika.hubova@gmail.com}} \& Ond\v rej Pejcha
\\
Institute of Theoretical Physics, Faculty of Mathematics and Physics, Charles University, Prague, Czech Republic\\
}

\date{Accepted XXX. Received YYY; in original form ZZZ}

\pubyear{2019}

\begin{document}
\label{firstpage}
\pagerange{\pageref{firstpage}--\pageref{lastpage}}
\maketitle

\begin{abstract}
We investigate kinematics of mass loss from the vicinity of the second Lagrange point L2 with applications to merging binary stars, common envelope evolution and the associated transient brightenings. For ballistic particle trajectories, we characterize initial velocities and positional offsets from L2 which lead to unbound outflow, fall back followed by a formation of a decretion disk, collision with the binary surface, or a hydrodynamic shock close to the binary, where some particle trajectories loop and self-intersect. The latter two final states occur only when the trajectories are initiated with offset from L2 or with velocity vector different from corotation with the binary. We find that competition between the time-dependent and steeply radially decreasing tidal torques from the binary, Coriolis force and initial conditions lead to a non-trivial distribution of outcomes in the vicinity of L2. Specifically, even for initial velocities slower than corotation, we find that a set of initial position offsets lead to unbound outflows. Our results will aid in the interpretation of the morphology of mass loss streams in hydrodynamic simulations.
\end{abstract}




\section{Introduction}

\begin{figure*}
	\includegraphics[width=0.8\textwidth]{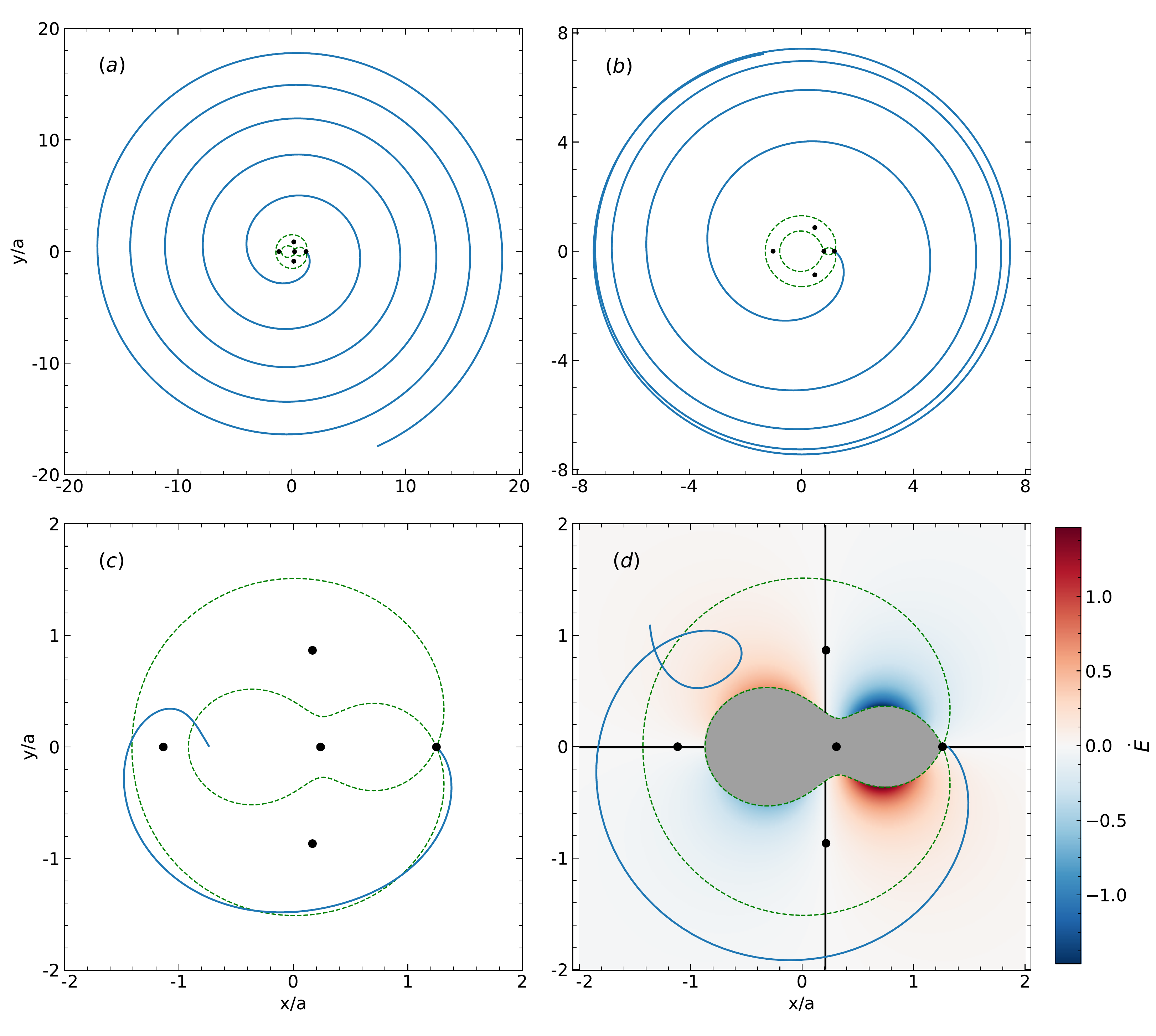}
    \caption{Examples of different types of trajectories (blue lines) in relation to the Roche equipotential passing through \ltwo (dashed green lines) and Lagrange points (filled black circles): (\emph{a}) thin equatorial outflow ($q=0.5$, $\mathbf{r}_0 = (0,0)$, $\mathbf{v}_0 = (0,0)$), (\emph{b}) decretion disk ($q=0.02$, $\mathbf{r}_0 = (0,0)$, $\mathbf{v}_0 = (0,0)$), (\emph{c}) collision with the binary ($q=0.5$, $\mathbf{r}_0 = (0, 0)$, $\mathbf{v}_0 = (0.5,-0.5)$)\deleted{,} and (\emph{d}) self-intersection of the stream ($q=0.4$, $\mathbf{r}_0 = ( 0.05, 0)$, $\mathbf{v}_0 = (0.6, -0.6)$). Panel (\emph{d}) also shows the instantaneous change of energy $\dot{E}$ in units of $a^2\omega^3$ as a function of position for particles initially in corotation with the binary orbit. Black lines separate regions of positive and negative $\dot{E}$. We did not evaluate $\dot{E}$ within the inner \ltwo equipotential, which we indicate in grey. The binary rotates counter-clockwise.}
    \label{fig:trajectories}
\end{figure*}

The evolution of binary stars, including exoplanets and their host stars, can be significantly influenced by the mass loss from the system. When the mass loss from a binary with semi-major axis $a$ and total mass $M$ occurs with velocities comparable to or lower than the binary escape velocity $\vesc = \sqrt{2GM/a}$ the morphology of the outflow is significantly affected by the gravitational forces of the system. In particular, the outer critical Lagrange point L2 plays a prominent role. The \ltwo point is the saddle point of the Roche potential located on the binary axis just outside the less massive component (Fig.~\ref{fig:trajectories}). At \ltwo, the gravitational and centrifugal accelerations in the corotating reference frame cancel out. This means that a only a small perturbation is required to eject an initially corotating test particle. Typical situations where this might arise are gradual expansion of the contact binary envelope, which was studied by \citet[]{webbink76,webbink77} for low-mass stars and \citet[]{flannery77} for high-mass stars. Mass loss from \ltwo can also result from rapid mass transfer between the binary components \citep[e.g.,][]{livio79,macleod18a}. In many cases, mass and angular momentum loss from L2 causes rapid shrinking of the orbit so that the binary soon experiences a common envelope evolution episode and might even merge into a single object \citep[e.g.,][]{paczynski76}. 

Recently, \citet{pejcha14} and \citet{pejcha17} argued that L2 mass loss lasting many hundreds and thousands of orbital periods was responsible for the pre-merger behaviour of V1309~Sco -- a contact binary that exhibited accelerating orbital period decay accompanied by changes in the light curve shape and overall brightening preceding its luminous red nova outburst \citep{tylenda11}. L2 mass loss or a similar process could lead to an equatorially-concentrated mass distribution that is later overtaken by a faster and more spherical explosion as is implicated in many supernovae, classical novae, luminous blue outbursts, and other transients \citep[e.g.,][]{li17,andrews18,smith18}. Furthermore, mass loss from the vicinity of \ltwo can feed the circumbinary disk of microquasar SS433 \citep{fabrika93} and may significantly affect the gravitational wave frequency and amplitude of extreme mass-ratio inspirals \citep{linial17}. 

Material initially positioned in corotation at L2 is always bound. However, after leaving L2 the time-changing gravitational field (i.e., binary tidal torques) transfer angular momentum and energy from the binary orbit to the L2 outflow \citep{kuiper41}. When considering purely ballistic motions of test particles, \citet{shu79} found that the final fate of L2 mass loss depends only on the binary mass ratio $q\le 1$. For $0.064 \le q \le 0.78$, the tidal torqueing is efficient enough that the L2 outflow achieves positive energy and leaves the binary to infinity with asymptotic velocities $v\lesssim 0.25\vesc$. For $q < 0.064$, tidal torqueing is inefficient owing to the small mass of the secondary component, and for $q > 0.78$, the initial energy of test particles corotating at L2 is too negative and the particle remains bound. For such a $q$, the L2 ballistic outflow does not achieve positive energy, falls back to the binary and establishes a viscous decretion disk. The picture becomes more complex when the ballistic particles are replaced by a realistic gaseous medium with shocks and radiative processes such as diffusion and cooling. \citet{pejcha16a,pejcha16b} found that, depending on the radiative cooling efficiency and the initial temperature, L2 mass loss leads to an equatorial outflow with a narrow or wide opening angle, a nearly isotropic outflow, a decretion disk or an inflated envelope.

Despite these advances, most of the work focusing specifically on L2 mass loss has so far been done assuming corotation and point-like injection at L2. This assumption must eventually become invalid when the timescale of orbital decay becomes shorter than the timescale for maintaining stellar rotation synchronized with the orbit\footnote{Since corotation is easier to maintain only in a relatively thin surface layer, the synchronization time-scale relevant for L2 mass loss can be much shorter than the tidal synchronization time-scale of the entire star.}, as is likely the case for common envelope evolution and stellar mergers \citep[e.g.,][]{paczynski76,meyer79}. Recently, \citet{macleod18b} studied the hydrodynamics of Roche-lobe overflow of a giant star and the associated mass loss from the vicinity of L2. For their binary mass ratio $q=0.3$, the L2 outflow should become unbound in ballistic calculations. However, \citet{macleod18b} found that the outflow actually remains bound and forms a decretion disk\deleted{,} because the initial tangential velocities of the stream are on average slower than corotation. This implies that the kinematics of the L2 mass loss stream is potentially sensitive to the actual initial conditions near L2. 

In this paper, we extend the previous ballistic calculations and investigate the trajectories and asymptotic properties of mass loss launched from a range of spatial positions around L2 and with a varying direction and magnitude of the initial velocity. In Section~\ref{sec:setting_methods}, we describe our setup for integrating ballistic particle trajectories, which we supplement for selected cases with smoothed particle hydrodynamics. In Section~\ref{sec:results}, we identify regions of positive and negative stream energy, stream collisions with the binary and loops leading to self-collisions within the stream. In Section~\ref{sec:discussion}, we summarize our findings and discuss implications for catastrophic binary interactions.

\section{Methods}
\label{sec:setting_methods}

Here we describe the geometry of the Roche potential and our numerical setup for integrating ballistic particle trajectories (Section~\ref{sec:motion}) as well as our implementation of smoothed particle hydrodynamics, which we use for investigation of hydrodynamical shocks (Section~\ref{sec:sph}).

\begin{figure*}
\includegraphics[width=1.0\textwidth]{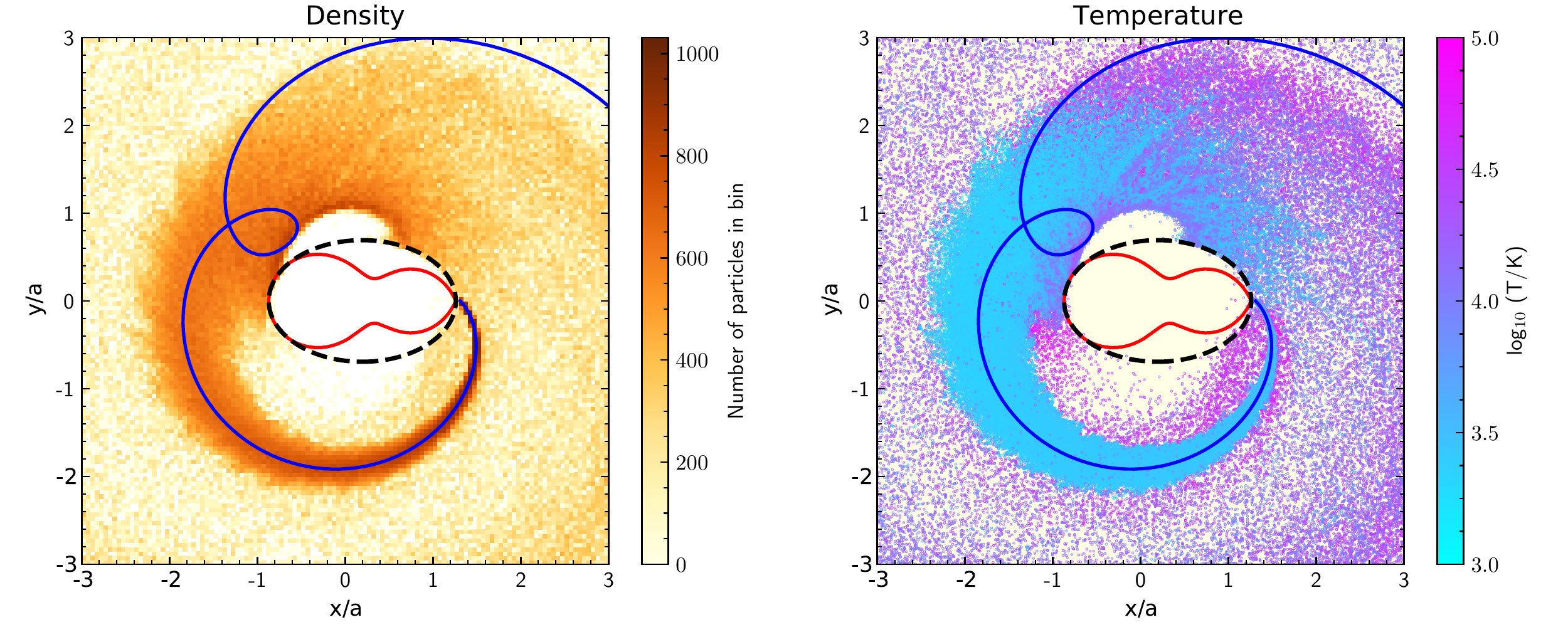}
\caption{Smoothed particle hydrodynamics of the stream based on the initial conditions from Fig.~\ref{fig:trajectories}, panel (d), which result in a loop and self-intersection of the ballistic trajectory (solid blue lines). The mass ratio of the system is $q=0.4$. The left panel shows the number density of equal-mass particles and the right panel shows the temperatures of individual particles. In both panels, particles are projected on to the binary orbital plane. The inner inflow boundary condition is a prolate ellipsoid (black dashed line), which approximates the outer critical surface passing through \ltwo (red solid line). The physical semi-major axis is $a=0.03$\,AU, the total binary mass is $M=1.4\,\msun$ and the binary rotates counter-clockwise. The calculation was set up by continuously injecting particles at a rate of $10^5$ particles per orbital period. The mass-loss rate is $10^{-3}\,\msun$\,yr$^{-1}$. The snapshot shown here is after the evolution has proceeded for about $7$ binary orbital periods.}
\label{fig:sph}
\end{figure*}

\subsection{Ballistic trajectories}
\label{sec:motion}

We study the restricted three-body problem in the non-inertial reference frame corotating with the binary circular orbit with angular frequency $\omega$. The positions of the two point-mass stars are fixed and the test particles launched from around L2 do not affect the gravitational field around the binary. The binary components have masses $M_1$ and $M_2$ and are positioned on the $x$-axis with barycentre at $x=0$ and the less massive star $M_2$ with $x>0$. We restrict our analysis to trajectories in the binary orbital plane $(x,y)$, because the problem has mirror symmetry with respect to the orbital plane and the motions in the vertical direction are trivial. Using $a$ as the unit of length, $a\omega$ as the unit of velocity, the net potential $\phi$ in units of $a^2\omega^2$ for the particle motion in the frame corotating with the binary is
\begin{equation}
\label{eq:roche_potential}
\phi = -\frac{\mu}{\sqrt[]{(x-1+\mu)^2+y^2}} - \frac{1-\mu}{\sqrt[]{(x+\mu)^2+y^2}} - \frac{1}{2}\left(x^2+y^2\right),
\end{equation}
where $\mu \equiv M_2/(M_1+M_2)$. 
The L2 point is located at $\xltwo > 0$ defined by solving 
\begin{equation}
\label{eq:position_of_l2_point}
f\xltwo - \frac{\mu}{\left(\xltwo-1+\mu\right)^2} -\frac{( 1-\mu)}{\left(\xltwo+\mu\right)^2} = 0.
\end{equation}
The parameter $f={\omega^\prime} / \omega$ is introduced to define an equilibrium point analogous to L2 (zero net of gravitational and centrifugal accelerations) in situations when the initial particle angular frequency $\omega'$ differs from $\omega$ \citep[see][for more details on Lagrange points in eccentric, non-corotationg orbits]{sepinsky07}. Unless explicitly stated otherwise, $f=1$ throughout the paper.

The particle trajectory is obtained by solving the equations of motion
\begin{subequations}
\begin{eqnarray}
\ddot{x} &=& -\pdv{\phi}{x} + 2\dot{y},\\
 \ddot{y} &=& -\pdv{\phi}{y} - 2\dot{x},
\label{eq:equations_of_motion}
\end{eqnarray}
\end{subequations}
using the \emph{scipy.integrate.solve\_ivp} function from the \emph{SciPy} library for \emph{Python}. We applied the Dormand-Prince method of order 4/5 (\emph{RK45}), which is an explicit method of the Runge-Kutta family with adaptive stepsize. During our computations, relative and absolute tolerances controlling the accuracy of $x$ and $y$ in units of $a$ were set to $10^{-12}$. 

We explored the trajectories for a range of initial positions $\mathbf{r}_0 = (x_0,y_0)$ and velocities $\mathbf{v}_0 = (\dot{x}_0,\dot{y}_0)$. For the sake of brevity, we express $\mathbf{r}_0$ relative to the L2 point so that particles starting in corotation at L2 have $\mathbf{r}_0 = (0,0)$ and $\mathbf{v}_0 = (0,0)$. The integration was terminated when one of the following events occurred. \emph{(i)} The particle crossed the inner equipotential line passing through the L2 point. We approximated the equipotential with a polygon and tested whether the trajectory entered inside every $0.25/\omega$ time units. \emph{(ii)} The trajectory intersected with itself. We determined this by approximating the trajectory with line segments and checking whether a newly-calculated segment intersected any of the previous ones. \emph{(iii)} The distance of the particle from system's centre of mass has exceeded $200a$. At this distance, the tidal torque from the central binary is sufficiently weak to not significantly change the asymptotic outcome.

\begin{figure*}
	\includegraphics[width=0.9\textwidth]{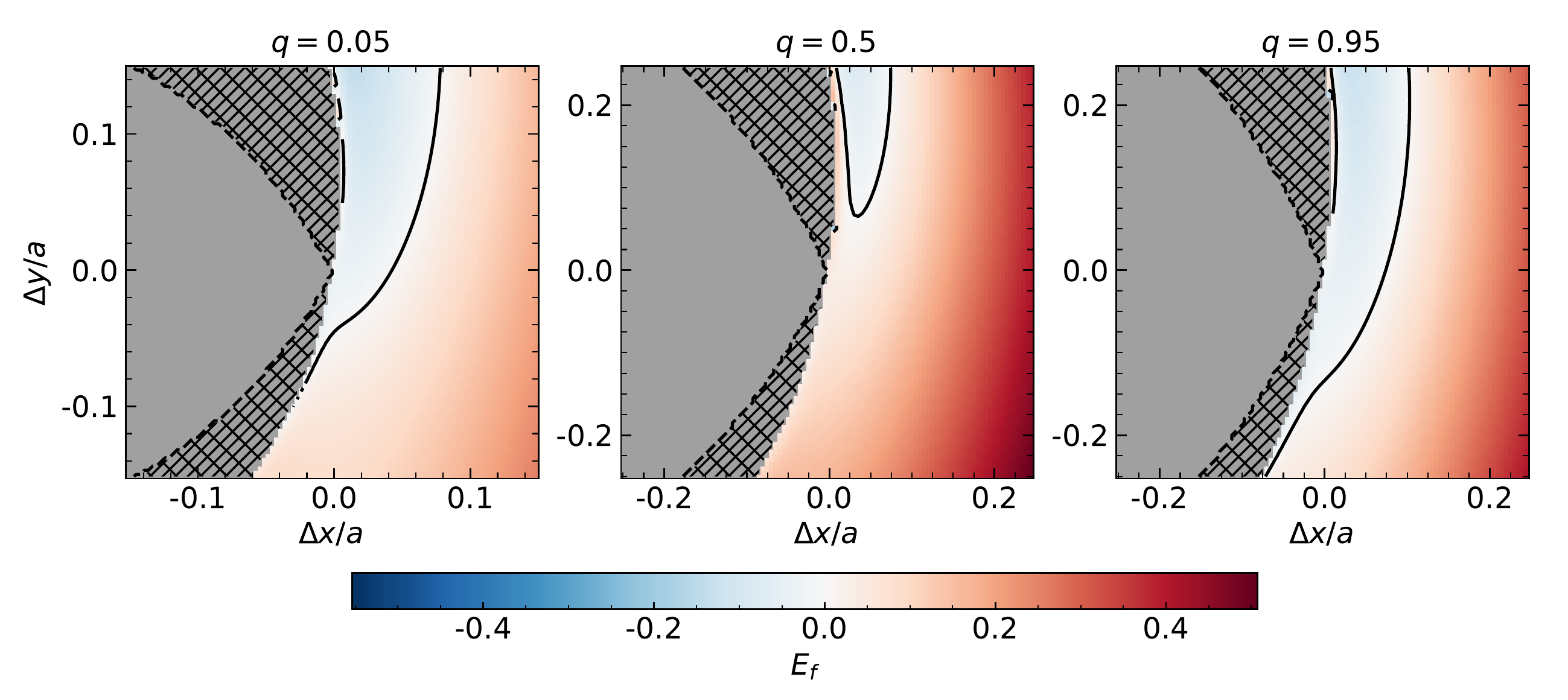}
    \caption{Final energy $\ef$ in units of $a^2\omega^2$ for particles injected in corotation with the binary orbit as a function of offset $(\Delta x, \Delta y)$ from the L2 point. Blue and red colors correspond to the final energy and black lines mark $\ef =0$. Hatched areas represent initial positions which result in collisions with the binary star. The binary star is marked in grey. }
    \label{fig:Ef_corotation}
\end{figure*}

\begin{figure*}
\includegraphics[width=0.67\textwidth]{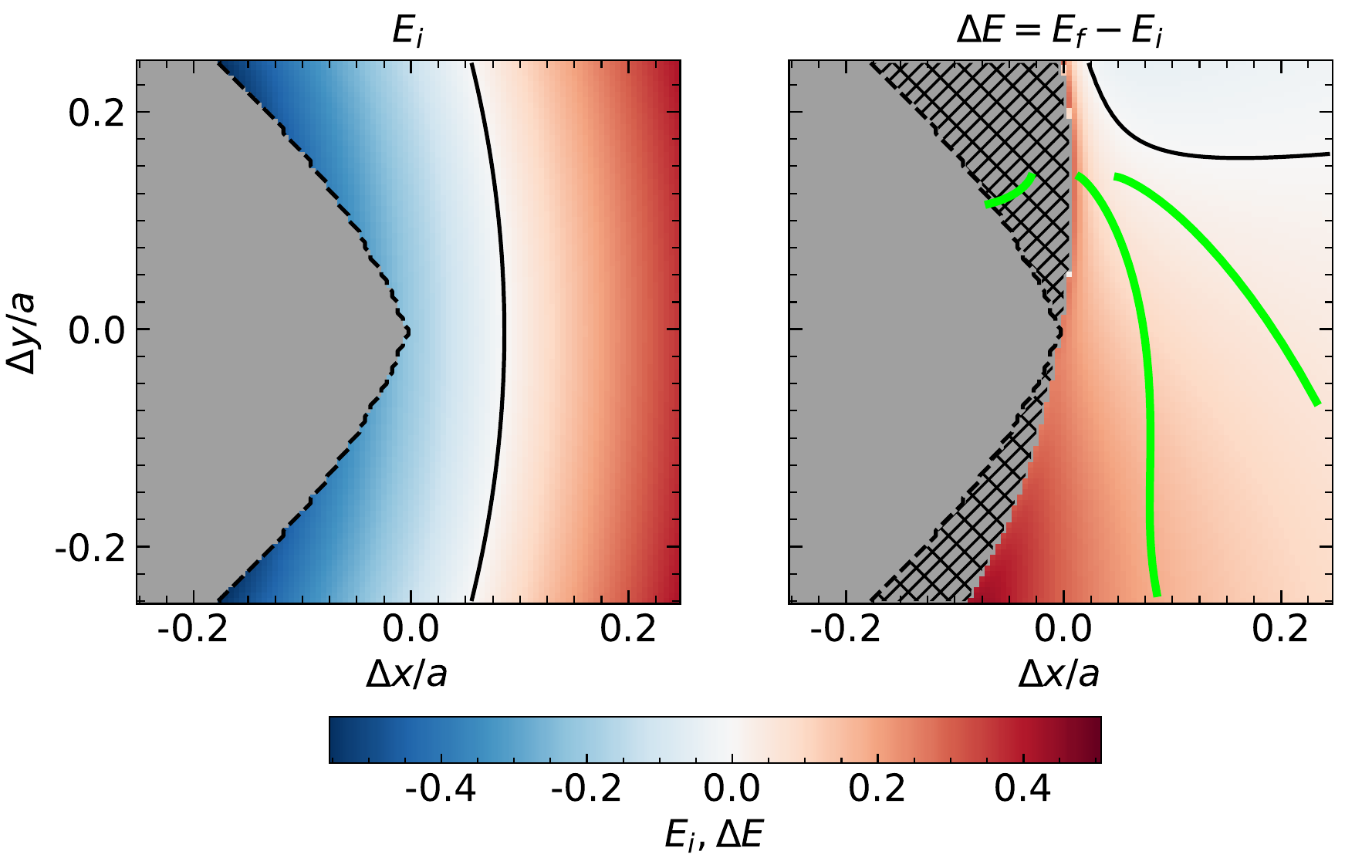}
\caption{Initial energy $\ei$ and energy gain $\Delta E=\ef-\ei$ in units of $a^2\omega^2$ as a function of offset $(\Delta x, \Delta y)$ from L2 for $q=0.5$. Black lines mark the locations of $\ei = 0$ and $\Delta E =0$, respectively. Green lines mark excerpts of ballistic trajectories started from three nearby positions but with different final outcomes.}
\label{fig:ei}
\end{figure*}

The energy $E$ and angular momentum $J$ per unit mass measured with respect to a stationary inertial observer can be computed from the position $(x,y)$ and the velocity $(\dot{x},\dot{y})$ of the particle in the corotating frame using the relations
\begin{eqnarray}
E &=& \phi + \frac{1}{2}\left(\dot{x}^2 + \dot{y}^2\right) + x^2 + y^2 +x\dot{y} -\dot{x}y,
\label{eq:energy}\\
J &=& x^2 + y^2 +x\dot{y} - \dot{x}y.
\label{eq:angular_momentum}
\end{eqnarray}
The specific energy computed from the above-mentioned expression has units $a^2\omega^2$, the specific angular momentum is given in units $a^2\omega$.
We denote initial and final energies of particles by $\ei$ and $\ef$, respectively. Along the particle's trajectory the value of Jacobi constant
\begin{equation}
\label{eq:jacobi_constant}
C =  \phi + \frac{1}{2}\left(\dot{x}^2+\dot{y}^2\right)
\end{equation}
is conserved. We see from equations (\ref{eq:energy}) and (\ref{eq:angular_momentum}) that along the particle's trajectory, the difference between its energy and angular momentum remains constant because $C = E - J$. Consequently, we focus on the behaviour of $E$, because $J$ can always be recovered from $C$.

\subsection{Smoothed particle hydrodynamics}
\label{sec:sph}

\begin{figure*}
	\includegraphics[width=0.8\textwidth]{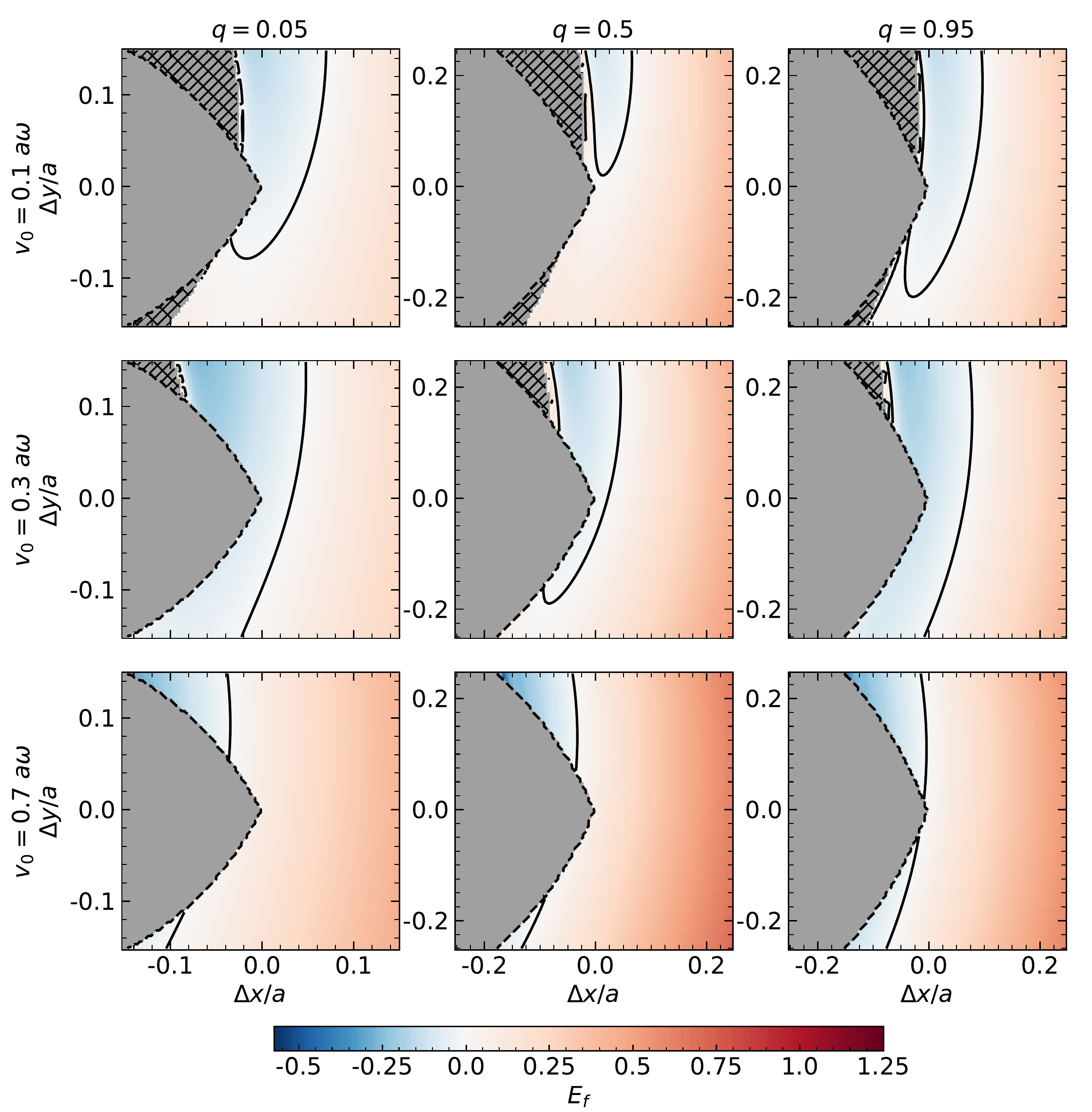}
    \caption{Final energy $\ef$ in units of $a^2\omega^2$ for particles ejected from the vicinity of the L2 point with initial velocity in radial direction and with magnitude $v_0$. The meaning of the symbols and lines is the same as in Fig.~\ref{fig:Ef_corotation}, except the colourbar, which covers wider range of $\ef$.}
    \label{fig:Ef_radial}
\end{figure*}

\begin{figure*}
	\includegraphics[width=0.8\textwidth]{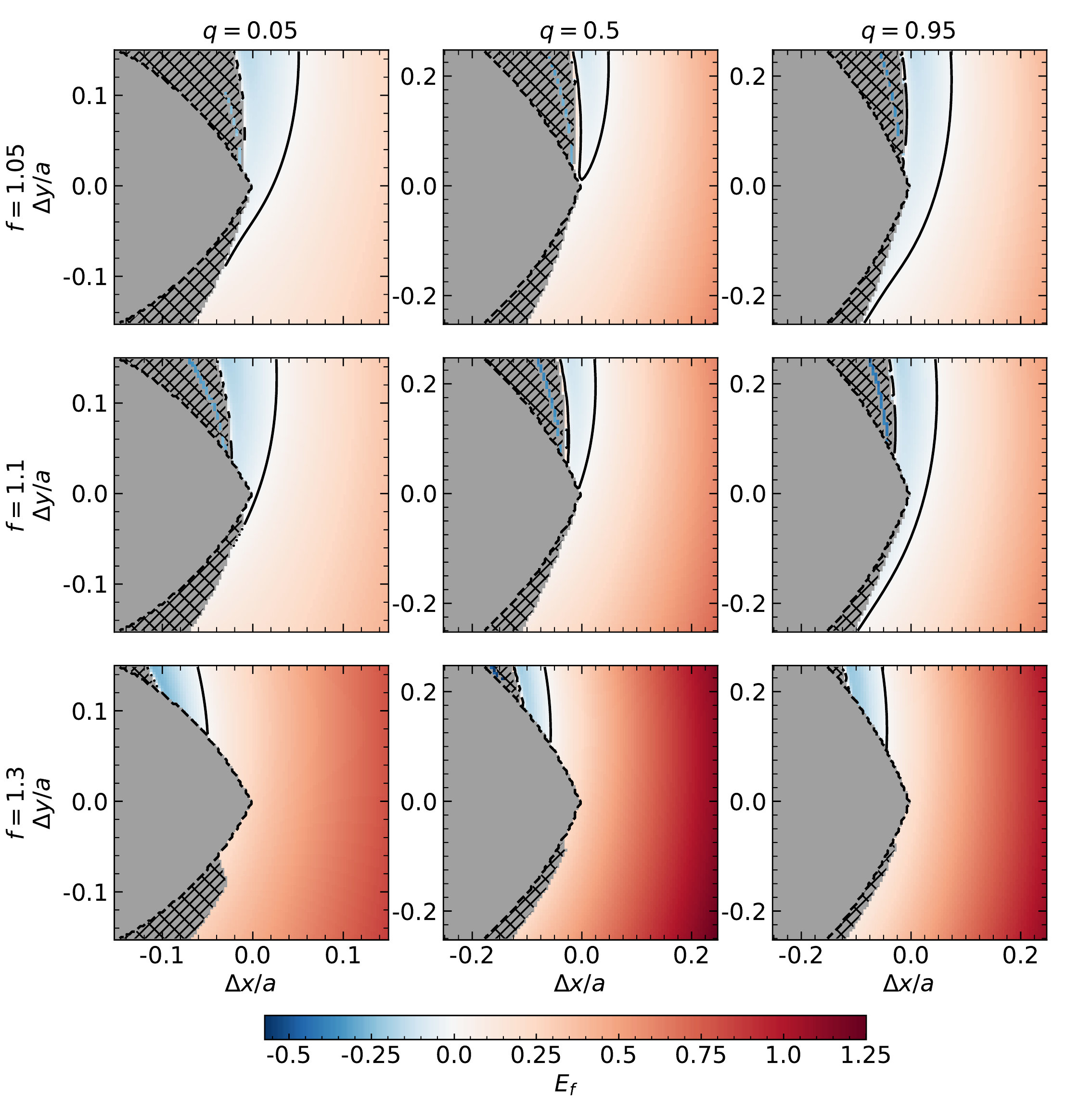}
    \caption{Final energy $\ef$ in units of $a^2\omega^2$ for particles ejected from the vicinity of the L2 point with initial angular frequency higher than corotation, as parameterized by $f$.  The meaning of symbols and lines is the same as in Fig.~\ref{fig:Ef_radial}.}       
   \label{fig:Ef_tangential,positive}
\end{figure*}

\begin{figure*}
	\includegraphics[width=0.8\textwidth]{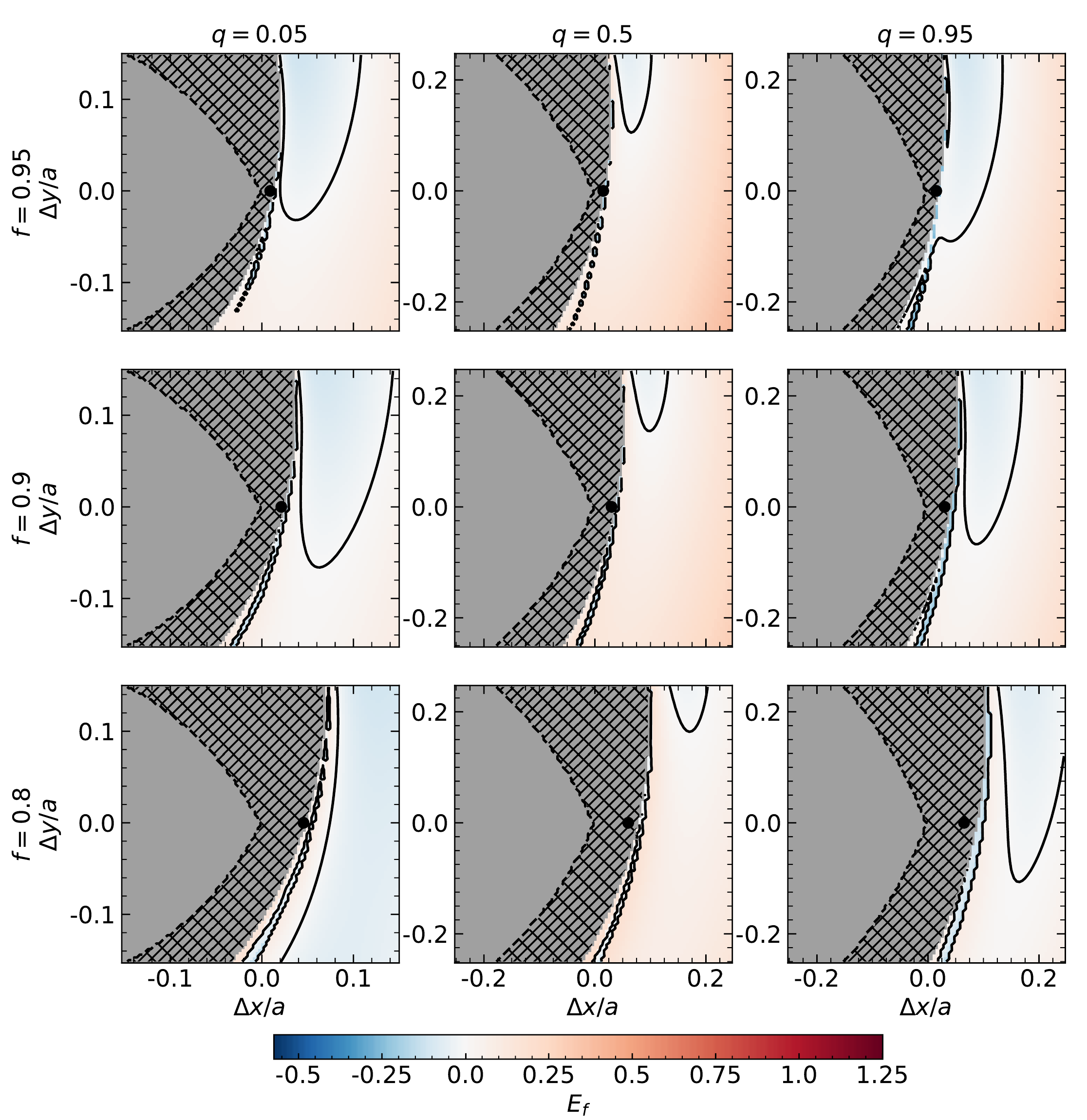}
    \caption{Final energy $\ef$ in units of $a^2\omega^2$ for particles ejected from the vicinity of the L2 point with initial angular frequency smaller than corotation, $f < 1$. The meaning of symbols and lines is the same as in Fig.~\ref{fig:Ef_radial}, but we also mark the modified position of L2 (Eq.~(\ref{eq:position_of_l2_point}), solid black circles).}
   \label{fig:Ef_tangential,negative}
\end{figure*}

Self-intersections of ballistic particle trajectories lead to hydrodynamic shocks. To investigate the gas dynamics in these cases, we utilize our smoothed particle hydrodynamics (SPH) code \citep{pejcha16a,pejcha16b,pejcha17}, which injects particles in a small volume around L2 and follows their dynamics in the external gravitational field of the binary star, approximated as two point masses. There is no gravitational interaction between the gas particles. Shocks are captured with standard artificial viscosity \citep{monaghan83,balsara95}. Radiative diffusion, cooling and irradiation by the central binary are not included. We employ the realistic equation of state of \citet{tomida13}. 

To approximately take into account re-accretion of particles by the binary star, we introduce an absorbing inner boundary. The natural choice for the inner boundary would be the outer critical surface of the Roche potential going through the \ltwo point. However, it would be too time-consuming to evaluate whether each particle crossed inside this three-dimensional implicitly-defined surface at every timestep. Instead, we approximate this surface with a prolate ellipsoid that exactly touches \ltwo and the corresponding opposite point on the $x$ axis (Fig.~\ref{fig:sph}). The other two axes are manually set to approximate the shape of the Roche equipotential near $M_1$. Although a proper way of implementing the inner boundary takes into account the properties of the  stellar surface, we simply remove any particle that crosses inside the ellipsoid. As a result, we cannot reliably model interaction of the mass loss with the star and our results are thus only indicative of a more complete calculation.

\section{Results}
\label{sec:results}

In this section, we describe the types of trajectories we identified in our work (Section~\ref{sec:trajectories}) followed by investigation of how these trajectory types depend on the initial position (Section~\ref{sec:position}) and velocity (Section~\ref{sec:velocity}).

\subsection{Types of particle trajectories}
\label{sec:trajectories}

In Fig.~\ref{fig:trajectories}, we show four types of particle trajectories identified in our work. The upper row shows previously known trajectory types, which were investigated in both ballistic and hydrodynamic approximations \citep[e.g.][]{shu79,pejcha16a,pejcha16b}. In panel (\emph{a}), the particle stream becomes unbound and forms a thin equatorial outflow heated by relatively weak internal spiral shocks. In panel (\emph{b}), the particles travel to a certain maximum distance, fall back to the binary, and eventually form a decretion disk or a more complicated object\footnote{For trajectories initially in corotation at L2, we find that the upper dividing line between bound and unbound trajectories is $q=0.792$, while \citet{shu79} found $q=0.78$. We were not able to trace the origin of this discrepancy.}. We detect this outcome as self-intersecting trajectories which occur with $E<0$. The bottom row shows trajectories newly identified in this work, which occur for more general initial conditions. Panel (\emph{c}) shows a particle colliding with the binary star. This results in hydrodynamical shocks and at least partial re-accretion of the gas stream. The nature of this interaction depends on the detailed structure of the binary star and it is thus difficult to assess the fractions of re-accreted and ejected material. Finally, panel (\emph{d}) shows a trajectory which loops and intersects with itself near the binary. Most of the loops occur with $E<0$ and we thus do not distinguish them from decretion disk formation. The trajectory in panel (\emph{d}) is an exception, where the loop occurs with $E>0$. This required an initial velocity vector in a particular direction and of relatively high magnitude. We do not find similar cases in the results shown below. We expect that trajectory loops lead to formation of hydrodynamical shocks in the stream and outside of the binary.

We used our SPH code to model the self-intersecting trajectory from panel (\emph{d}) of Fig.~\ref{fig:trajectories}. We show a representative snapshot of the final steady-state configuration in Fig.~\ref{fig:sph}. After the simulation begins, the centre of the SPH particle stream follows the ballistic stream trajectory. As the stream turns back to the binary, around the third Lagrange point, some particles cross the inner boundary and are removed from the simulation without causing a hydrodynamic shock\footnote{When we made the inner absorbing boundary smaller, these particles formed a narrow accretion funnel similar to what is seen in accreting circumbinary disks \citep[e.g.][]{artymowicz96,munoz16}.}. As the stream centre gets to the turning point of the ballistic particle loop, SPH particles collide and the inner rim of the stream becomes shocked. This is visible as a dense and hot arc immediately below and to the right of the ballistic particle loop in Fig.~\ref{fig:sph}. However, part of the shocked rim is now on the leading side of the binary, loses energy and angular momentum from the tidal torques and soon disappears below the inner boundary. However, some of the shock-heated gas also expands away from the orbital plane and fills the binary surroundings with a hot medium. Some fraction of the gas becomes unbound and flows away. To summarize, self-intersecting trajectories deposit gas with low angular momentum and/or high temperature in the vicinity of the binary star. This is somewhat similar to decretion disk formation although, for a decretion disk and moderate $q$, the shock is farther away from the binary.

\subsection{Dependence on initial position}
\label{sec:position}

In Fig.~\ref{fig:Ef_corotation}, we show the trajectory type and final energy as a function of the starting position $\mathbf{r}_0 = (\Delta x, \Delta y)$, relative to \ltwo, and the binary mass ratio. All trajectories were initiated with $\mathbf{v}_0 = (0,0)$. Trajectories launched very close to the binary surface end up eventually colliding with the binary (grey hatched areas). The hatched area on the $\Delta y > 0$ side is larger than for $\Delta y < 0$, because trajectories need to pass around the physical barrier near L2, where the star extends to largest $x$. Trajectories launched further away from the binary can escape to infinity ($\ef > 0$) or return to the binary and form a decretion disk ($\ef<0$).  We see in Fig.~\ref{fig:Ef_corotation} that negative energies always occupy a lobe-like region, which is either confined to $\Delta y>0$, for intermediate $q$, or extends to L2 and $\Delta y<0$, for small and high $q$. Interestingly, there is always a narrow region of $\ef>0$ located at $\Delta y > 0$ and adjacent to initial conditions that lead to collisions with the binary.

To better understand these features, we show in Fig.~\ref{fig:ei} the initial energy $\ei$ (left panel) and the energy gain $\Delta E = \ef-\ei$ (right panel) as a function of positional offset. We also show excerpts of three trajectories with different starting points. Furthermore, Fig.~\ref{fig:trajectories}\emph{d} shows time derivative of $E$ as a function of position for particles initially in corotation with the binary. In Fig.~\ref{fig:ei}, $\ei$ is higher away from the binary, because both potential and kinetic energy increase. The potential energy increases, because the particle is farther away from the two point masses. The kinetic energy increases, because higher velocities are needed to ensure corotation with the binary. The initial energy is symmetric with respect to the $x$ axis. The energy gain is set by the efficiency of tidal torquing, which is illustrated in Fig.~\ref{fig:trajectories}\emph{d}. In the vicinity of L2 tidal torques increase $E$ for particles lagging behind ($\Delta y <0$) and decrease $E$ for particles leading ahead ($\Delta y>0$). We also see that there is always a positive energy gain adjacent to the region of collisions with the binary surface, even for $\Delta y > 0$. As is illustrated by the middle particle trajectory (Fig.~\ref{fig:ei}, green line), trajectories launched from this region pass around the obstacle of the binary surface as they are accelerated by the Coriolis force and Roche potential gradient. After crossing to $\Delta y<0$, the trajectory is still very close to the binary and experiences strong tidal torqueing (Fig.~\ref{fig:trajectories}\emph{d}). The energy gain in this region more than compensates for $\ei < 0$ and for energy losses while the particle was leading ahead of the binary.

\subsection{Dependence on initial velocity}
\label{sec:velocity}

We now investigate changes to the picture outlined in Section~\ref{sec:position} when the trajectory starts with non-zero velocity in the corotating frame. It is instructive to first investigate purely radial initial velocities, for which we show the results in Fig.~\ref{fig:Ef_radial}. As the velocity in the radial direction increases, the trajectories reach farther away from the binary. As a result, the area of initial conditions leading to collisions with the binary surface shrinks. At the same time, for small initial velocities, $|\mathbf{v}_0| \equiv v_0 \lesssim 0.3$, the area with  $\ef<0$ increases. This occurs because the efficiency of tidal torqueing sharply drops with the distance from the binary (Fig.~\ref{fig:trajectories}\emph{d}) and the trajectories do not gain enough energy to become unbound despite higher $\ei$. For even higher $v_0$, the initial kinetic energy is so large that the area of $\ef>0$ increases again (bottom row of Fig.~\ref{fig:Ef_radial}).

An alternative way to view our results is to investigate the range of $q$ leading to $\ef>0$. We studied trajectories starting from $\mathbf{r}_0= (0,0)$ on a fine grid of $q$ and radial velocity magnitude $v_0$. As $v_0$ increases the range of $q$ leading to $\ef>0$ slightly increases and peaks around $v_0 \approx 0.04$. For higher $v_0$, the range shrinks and achieves a minimum around $v_0 \approx 0.3$, where $\ef>0$ is achieved only for $0.15 \lesssim q \lesssim 0.27$. For $v_0 \gtrsim 0.7$ all $q$ lead to $\ef>0$.

In Fig.~\ref{fig:Ef_tangential,positive}, we show $\ef$ for trajectories initially on a circular orbit but with $\omega' = f\omega$, $f>1$. The results are qualitatively similar to the radial velocity: for $f \lesssim 1.1$, it is easier to evade collision with the binary surface, especially for the initial conditions of the leading side of the binary motion, but often at the price of less efficient tidal torqueing, because the trajectories move further away from the binary. As a result, the contour of $\ef=0$ gets closer to \ltwo. For $f\gtrsim 1.1$, the high initial velocity results in an increase of the area with $\ef > 0$. We also note a thin line of initial positions with $\ef<0$ passing through the hatched area at $\Delta y >0$. These trajectories looped and self-intersected  before they collided with the binary (Fig.~\ref{fig:trajectories}d).

In Fig.~\ref{fig:Ef_tangential,negative}, we show the results for $f<1$. As expected, the area of initial conditions leading to collisions with the binary is significantly larger owing to lower kinetic energy. However, if we relax the assumption of corotation, the Roche potential is modified  and so are the  positions of Lagrange points \citep{sepinsky07}. Specifically, for  $f<1$, $\xltwo$ increases as we show in Fig.~\ref{fig:Ef_tangential,negative}. Binary stars with surface layers rotating slower than the orbital frequency would lose mass from the vicinity of the modified L2 point. In such cases, there still exist initial positions near the modified L2 which eventually achieve $\ef >0$, at least for moderate $q$. However, trajectory self-intersections (Fig.~\ref{fig:trajectories}d) also become prominent especially for smaller $f$. Furthermore, in generating Fig.~\ref{fig:Ef_tangential,negative}, we still checked for collisions with the binary surface assuming the original corotating equipotential. It is likely that taking into account the modified equipotential would increase the number of colliding trajectories.

Finally, we also investigated initial velocity vectors pointing to an arbitrary direction. In some cases, the final outcome of the trajectory sensitively depends on the initial conditions. This suggests a possibility of deterministic chaos. Typically, these features occur for trajectories closely approaching some of the other Lagrange points. Around these stationary points, the direction of the Roche potential gradient varies significantly and the trajectories effectively scatter to a wide range of directions. However, this behaviour would likely result in hydrodynamic shocks which remain to be explored in more detail in the future.

\section{Discussion and conclusions}
\label{sec:discussion}

We have performed detailed investigations of asymptotic behaviour of ballistic trajectories launched from the vicinity of the \ltwo point. In addition to previously known outcomes of unbound outflow and fallback forming a decretion disk, we identified a relatively narrow space of initial conditions where the ballistic trajectories self-intersect close to the binary star (Fig.~\ref{fig:trajectories}d). This leads to formation of hydrodynamic shocks, as we illustrate with smoothed particle hydrodynamics (Fig.~\ref{fig:sph}).

We studied how the final outcome depends on the initial position (Fig.~\ref{fig:Ef_corotation}) and velocity vector (Figs.~\ref{fig:Ef_radial}, \ref{fig:Ef_tangential,positive} and \ref{fig:Ef_tangential,negative}). We find that all types of outcomes are present in a relatively narrow extent of physical offsets from \ltwo. For example, there generally exists a thin arc of positive final energies on the leading side of the binary, where tidal torqueing is surprisingly efficient (Fig.~\ref{fig:ei}, right panel, and Fig.~\ref{fig:trajectories}\emph{d}). Trajectories initiated with a small offset from this arc either collide with the binary or do not achieve positive final energy. Furthermore, even for trajectories initiated about $20$\% slower than corotation, there still exist regions where the trajectories end up unbound.

Because a realistic stream leaving the binary from \ltwo would start from a range of initial positions and velocities, our results imply that interpreting hydrodynamical simulations of binary mass loss with ballistic trajectories needs to take into account the sensitivity to initial conditions. This agrees with findings of \citet{macleod18b}, who also point out that hydrodynamical interactions within the stream, which we cannot study with ballistic trajectories, contribute to the final outcome. However, our Fig.~\ref{fig:Ef_corotation} suggests that very small shifts in the initial $\Delta x$ of ballistic trajectories with $\Delta y > 0$, such as those used in Fig.~5 of \citet{macleod18b}, could potentially result either in collisions with the binary or even positive asymptotic energies. Furthermore, the mass loss from the vicinity of \ltwo is affected by binary surface perturbations owing to convective motions, meridional and zonal circulations, and magnetic fields, all of which intensify as the mass-loss rate runs away. These physical processes are not resolved in most of the binary hydrodynamical simulations.

Despite these uncertainties, our results already offer clues how to interpret observations of suspected \ltwo mass loss. In their study of luminous red nova V1309~Sco, \citet{pejcha17} found that the initial temperature of the \ltwo stream needs to dramatically increase in the last about 75 days to explain the observed upward curvature of luminosity. Their results imply that $f \approx 0.95$ at 75 days before the merger. Our results in Figure~\ref{fig:Ef_tangential,negative} show that even for $f=0.95$ there is an increased fraction of initial conditions that lead to collision with the binary surface and that $\ef$ of unbound particles is lower than for $f=1$. Although we do not know the distribution of initial positions of the stream, it is likely that the fraction of \ltwo stream remaining near the binary increases with decreasing $f$. This material will shock-heat to temperatures similar to the kinetic temperature of the binary, which could naturally explain the increase of initial temperature of the \ltwo stream.

\section*{Acknowledgements}

We thank our referee, Prof. Christopher Tout, for comments and suggestions, which improved the paper. Our research is supported by Horizon 2020 ERC Starting Grant ``Cat-In-hAT'' (grant agreement \#803158), award PRIMUS/SCI/17 from Charles University and INTER-EXCELLENCE grant LTAUSA18093 from the Czech Ministry of Education, Youth, and Sports.




\bibliographystyle{mnras}
\bibliography{bibliography} 

\bsp	

\end{document}